\documentclass[apjl,twocolumn]{emulateapj}
\usepackage{graphicx,natbib}

\shorttitle{Counterpart Identification with PACS imaging}
\shortauthors{Dannerbauer et al.}

\begin{document}


\title{Unveiling Far-Infrared Counterparts of Bright Submillimeter Galaxies Using PACS Imaging}


\author{H. Dannerbauer\altaffilmark{1},
\email{helmut.dannerbauer@cea.fr}
  E. Daddi\altaffilmark{1},
  G.~E. Morrison\altaffilmark{2,3},
  B. Altieri\altaffilmark{4},
  P. Andreani\altaffilmark{5,6},
  H. Aussel\altaffilmark{1},
  S. Berta\altaffilmark{7},
  A. Bongiovanni\altaffilmark{8,9},
  A. Cava\altaffilmark{8,9},
  J. Cepa\altaffilmark{8,9},
  A. Cimatti\altaffilmark{10},
  H. Dominguez\altaffilmark{11},
  D. Elbaz\altaffilmark{1},
  N. F{\"o}rster Schreiber\altaffilmark{7},
  R. Genzel\altaffilmark{7},
  C. Gruppioni\altaffilmark{11},
  B. Horeau\altaffilmark{1},
  H.S. Hwang\altaffilmark{1},
  E. Le Floc'h\altaffilmark{1},
  J. Le Pennec\altaffilmark{1},
  D. Lutz\altaffilmark{7},
  G. Magdis\altaffilmark{1},
  B. Magnelli\altaffilmark{7},
  R. Maiolino\altaffilmark{12},
  R. Nordon\altaffilmark{7},
  A.M. P{\'e}rez Garc{\'i}a\altaffilmark{8,9},
  A. Poglitsch\altaffilmark{7},
  P. Popesso\altaffilmark{7},
  F. Pozzi\altaffilmark{10},
  L. Riguccini\altaffilmark{1},
  G. Rodighiero\altaffilmark{13},
  A. Saintonge\altaffilmark{7},
  P. Santini\altaffilmark{12},
  M. Sanchez-Portal\altaffilmark{4},
  L. Shao\altaffilmark{7},
  E. Sturm\altaffilmark{7},
  L. Tacconi\altaffilmark{7},
  I. Valtchanov\altaffilmark{4}
}


\altaffiltext{1}{Laboratoire AIM, CEA/DSM - CNRS - Universit\'{e} Paris Diderot, DAPNIA/Service d'Astrophysique, CEA Saclay, Orme des Merisiers, F-91191 Gif-sur-Yvette Cedex, France}
\altaffiltext{2}{Institute for Astronomy, University of Hawaii, Manoa, Hawaii 96822, USA}
\altaffiltext{3}{Canada-France-Hawaii Telescope Corp., Kamuela, Hawaii 96743, USA}
\altaffiltext{4}{Herschel Science Centre}
\altaffiltext{5}{ESO, Karl-Schwarzschild-Str. 2, D-85748 Garching, Germany}
\altaffiltext{6}{INAF - Osservatorio Astronomico di Trieste, via Tiepolo 11, 34143
Trieste, Italy}
\altaffiltext{7}{Max-Planck-Institut f\"{u}r Extraterrestrische Physik (MPE), Postfach 1312, 85741 Garching, Germany}
\altaffiltext{8}{Instituto de Astrof{\'i}sica de Canarias, 38205 La Laguna, Spain}
\altaffiltext{9}{Departamento de Astrofísica, Universidad de La Laguna, Spain}
\altaffiltext{10}{Dipartimento di Astronomia, Universit{\`a} di Bologna, Via Ranzani 1,
40127 Bologna, Italy}
\altaffiltext{11}{INAF-Osservatorio Astronomico di Bologna, via Ranzani 1, I-40127 Bologna, Italy}
\altaffiltext{12}{INAF - Osservatorio Astronomico di Roma, via di Frascati 33, 00040 Monte Porzio Catone, Italy}
\altaffiltext{13}{Dipartimento di Astronomia, Universit{\`a} di Padova, Vicolo dell'Osservatorio 3,
35122 Padova, Italy}


\begin{abstract}
We present a search for Herschel-PACS counterparts of dust-obscured,
high-redshift objects previously selected at submillimeter and
millimeter wavelengths in the Great Observatories Origins Deep Survey
North field. We detect 22 of 56 submillimeter galaxies ($39\%$) with a
SNR of $\ge3$ at 100~$\mu$m down to 3.0~mJy, and/or at 160~$\mu$m down
to 5.7~mJy. The fraction of SMGs seen at 160~$\mu$m is higher than that at 100~$\mu$m. About 50\% of radio-identified SMGs are
associated with PACS sources.  We find a trend between the SCUBA/PACS
flux ratio and redshift, suggesting that these flux ratios could be
used as a coarse redshift indicator. PACS undetected submm/mm selected
sources tend to lie at higher redshifts than the PACS detected ones. A
total of 12 sources ($21\%$ of our SMG sample) remain unidentified and
the fact that they are blank fields at Herschel-PACS and VLA 20
cm wavelength may imply higher redshifts for them than for the
average SMG population (e.g., $z>3-4$). The Herschel-PACS imaging of
these dust-obscured starbursts at high-redshifts suggests that their
far-infrared spectral energy distributions have significantly
different shapes than template libraries of local infrared galaxies.
\end{abstract}


\keywords{Galaxies: high-redshift --- Galaxies: individual (GN13 alias HDF850.4) --- Galaxies: starburst --- Cosmology: observations --- Infrared: Galaxies --- Submillimeter: galaxies}




\section{Introduction}
Several hundred dust-enshrouded high-z sources have been selected
through submm/mm imaging with bolometer cameras like SCUBA, LABOCA,
AzTEC, MAMBO
\citep[e.g.,][]{hug98,dan02,dan04,sma02,cop06,pop06,ber07,per08,wei09}. The
large beam size in the (sub)millimeter (e.g., MAMBO: 11\arcsec; SCUBA:
15\arcsec; LABOCA: 19\arcsec) hampers the identification of these
so-called Submillimeter Galaxies \citep[SMGs; see for a
review][]{bla02} based on bolometer data only. The most
obvious choice for obtaining the subarcsecond accurate positions of
the dust continuum is (sub)millimeter interferometric continuum
observations \citep[e.g.,][]{dan02,dan08,you07}. However the slow
mapping speed of (sub)millimeter interferometers does not allow
us to study a large number of sources. The most suitable tool for
counterpart identification is interferometric observations at radio
wavelengths.  About $50\% - 80\%$ of these submm/mm sources have been
identified, mainly based on radio observations, and the peak of the
redshift distribution of the radio-identified SMG population lies at
$<z>=2.2$ \citep{cha05}. However, the effect of the bias toward
the true redshift distribution introduced by the radio selection
technique is still under debate, as SMGs even beyond $z=4$ have been
already detected by the VLA at 20~cm
\citep[e.g.,][]{dad09a,dad09b,cap08,schi08,cop09,knu10,mor10}. The sample of
unidentified SMGs could be either spurious sources or lie at extreme
redshifts but no systematic studies on this subsample of SMGs have
been conducted so far.

The launch of the Herschel observatory \citep{pil10} promises a 
different perspective of SMGs than provided by radio observations only
and offers a higher mapping speed than (sub)mm interferometric
observations. Deep PACS \citep{pog10} imaging at 100~$\mu$m and
160~$\mu$m on the Great Observatories Origins Deep Survey North
(GOODS-N) will sample the FIR emission of these dust-enshrouded high-z
objects and enable us to study in detail their far-infrared spectral
energy distribution (SED), redshift distribution, dust temperatures
\citep{mag10,elb10,chan10} and dust masses \citep{san10}. In absence
of interferometric observations at mm and cm wavelengths, the
Herschel-PACS beamsize at 100~$\mu$m (160~$\mu$m) of 6.7\arcsec\/
(11.0\arcsec\/) provide a more accurate location of the dust emission
than the bolometer data taken with SCUBA, AzTEC, MAMBO or LABOCA. \\
\newline In this letter we discuss our search for PACS counterparts at
100~$\mu$m and 160~$\mu$m, explore the diagnostic potential of
Herschel-PACS for the counterpart identification and compare it with
the widely used identification approach using VLA observations. In
comparison, the Spitzer-MIPS 70~$\mu$m and 160~$\mu$m imaging of the
GOODS North region performed by \citet{huy07} in the pre-Herschel era
detected at relatively high significance only two (one) out of 30 SMGs
at 70~$\mu$m (160~$\mu$m), at rather low redshifts ($z=0.5$ and
$z=1.2$). The FIR observations presented here will enable us to study
a significant sample of SMGs in the far-infrared wavelength regime.

\section{Far-Infrared Association of Submillimeter Galaxies}
PACS observations of the GOODS North region at 100~$\mu$m and
160~$\mu$m were taken during the Herschel Science Demonstration Phase
in autumn 2009 and are part of the Guarantee Time extragalactic PACS
survey 'PEP: The PACS Evolutionary Probe' (PI: D. Lutz). The final
images achieve $3\sigma$ sensitivities of $\sim 3.0$~mJy and $\sim
5.7$~mJy at 100~$\mu$m and 160~$\mu$m respectively. \citet{ber10} describe in
their appendix the Herschel-PACS observations, data reduction and the
blind extraction of PACS sources with SNR$\ge3$. Lutz et al. (in
prep.) will present the fluxes of PACS sources.  The PACS images are
aligned on the GOODS imaging products. Complementary to the Herschel PACS
observations, we use for this work data from the VLA at 1.4 GHz
\citep{mor10} and Spitzer-MIPS at 24~$\mu$m (Dickinson et al., in
prep.).
   \begin{figure}[!t]
   \centering
\includegraphics[width=8cm]{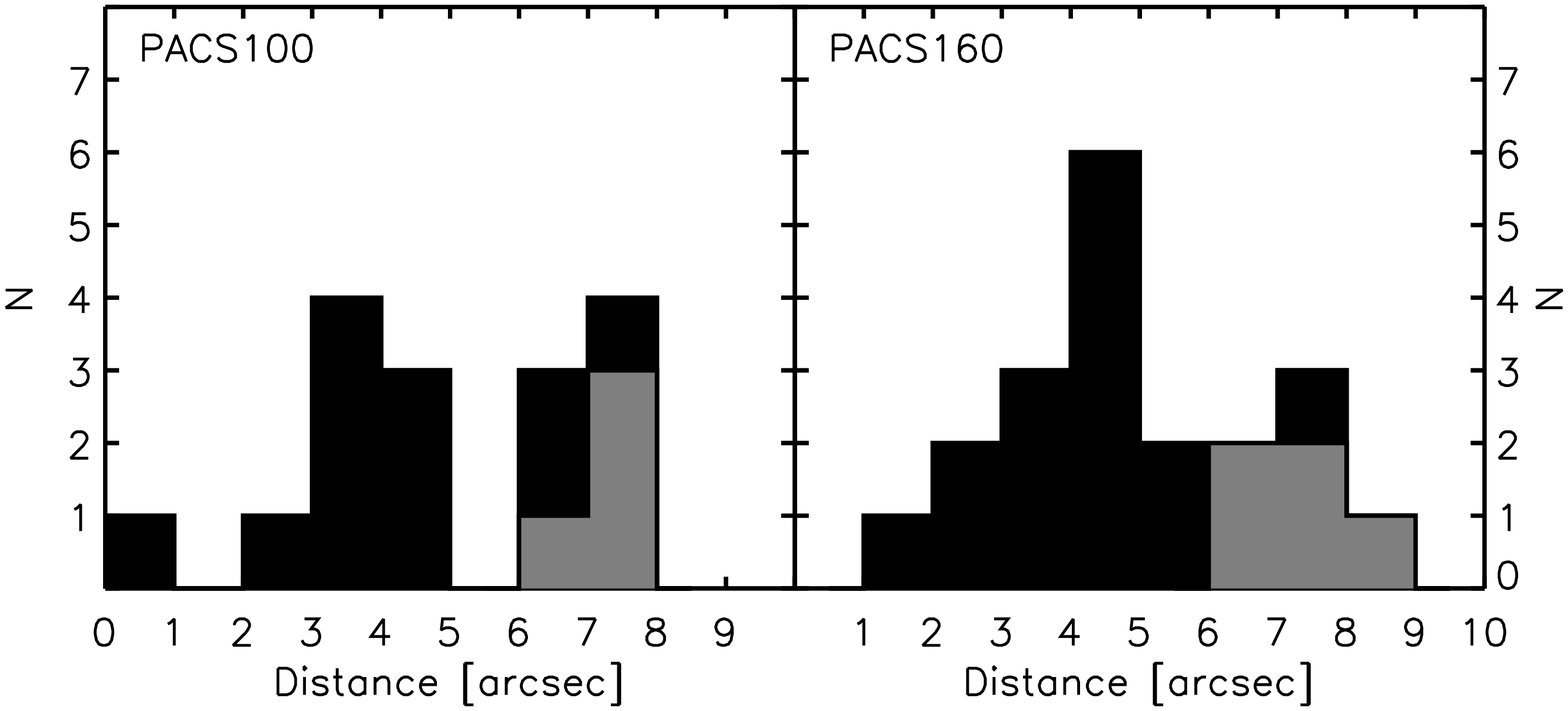}
\includegraphics[width=8cm]{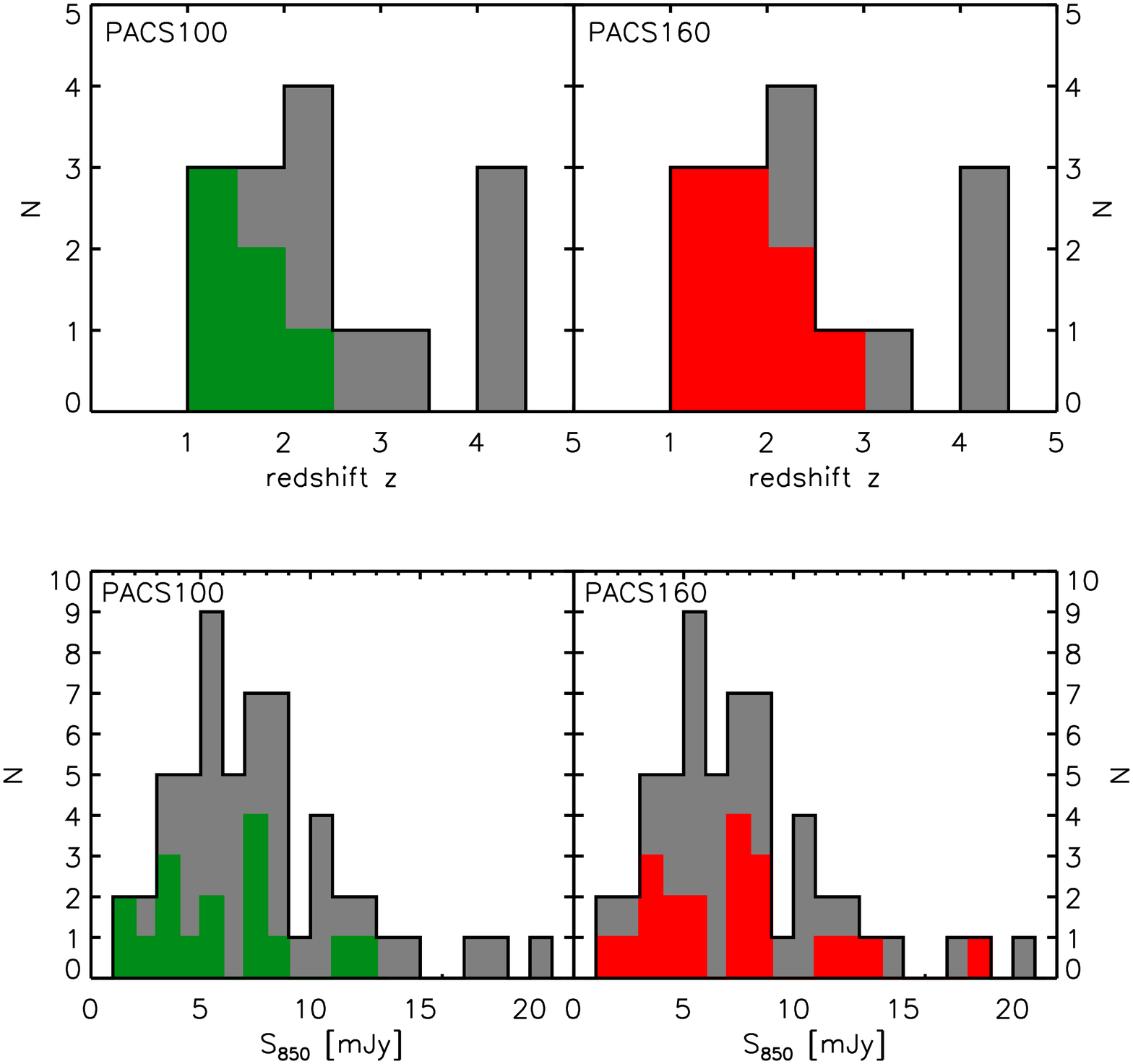}
      \caption{{\it Top panel:} Histogram of distance between the submm/mm position
      from the bolometric map and PACS positions of associated FIR
      sources at 100~$\mu$m (left) and 160~$\mu$m (right). Black and
      gray indicate secure and possible PACS counterparts. {\it Middle panel:} We show the redshift distribution for our SMG sample with spectroscopic redshifts (gray) and overplot our PACS detections at 100~$\mu$m (green) respectively 160~$\mu$m (red). {\it Lower panel:} We show the SCUBA flux distribution of our SMG sample (gray) and overplot our PACS detections at  100~$\mu$m (green) respectively 160~$\mu$m (red).}
         \label{fig:pacsdetection}
   \end{figure}
%

In the past years, several groups surveyed the GOODS North region
using the bolometer cameras SCUBA, AzTEC and MAMBO
\citep{hug98,bor03,pop05,wan04,per08,gre08}. These observations
discovered about 150 SMGs at 850~$\mu$m, 1.1 and 1.2~mm. Robustly
identified VLA and MIPS 24~$\mu$m counterparts are already known in
the literature for SCUBA and AzTEC sources \citep{pop06,cha09}. In
addition, \citet{gre08} presented VLA counterparts for 11 out of 30
MAMBO sources. 

We search for Herschel-PACS counterparts of SMGs that are either
detected with a SNR of $\geq4$ or detected by at least two different
surveys. 56 SMGs fulfill this criterion which should assure a robust
SMG sample to work with. Our sample\footnote{In case that an SMG is
detected at submm and mm bands, we give preference the SCUBA detection
and include it in our sample. For mm-only detected sources we list, if
available, the AzTEC otherwise the MAMBO detection.} consists of 36
SCUBA, 12 AzTEC and 8 MAMBO sources and for 15 SMGs spectroscopic
redshifts (SMG$_{spec}$) have been already obtained. We match our SMG
sample with the PACS 100~$\mu$m and 160~$\mu$m blind catalogue and search for
counterparts within a radius of 5.5\arcsec\/ for MAMBO sources,
7.5\arcsec\/ for SCUBA sources and 9\arcsec\/ for AzTEC sources. As a
sanity check of the blind catalogue, we inspected by eye the search
region in the PACS imaging. The search circles that we have applied
correspond to the beam size (FWHM) of the different bolometric
datasets and should guarantee that no reliable associations to SMGs
will be missed. We also searched the VLA 1.4 GHz map and the MIPS
24~$\mu$m images for counterparts at the SMG positions.

We calculate the corrected Poissonian probability {\it p} that an
association of SMGs within the search radius is a chance
coincidence. This approach \citep[see for details][]{dow86} corrects
the simple Poissonian probability of a detected association for the
possibility of associations of different nature but similar
probability. The derived probability of PACS and MIPS association is
based on raw number counts in GOODS North.  We search for VLA
counterparts of SMGs down to $3\sigma$. However, the VLA source
catalogue \citep{mor10} is only reliable down to 20~$\mu$Jy ($\sim5
\sigma$). Therefore, we assess the reliability of VLA counterparts
relying on published number counts \citep[e.g.,][]{fom06}. Similar to previous studies, we define following quality criteria for
assessing the robustness of identified candidate counterparts. We
classify association of SMGs with $p\leq0.05$ as secure and with
$0.05<p\leq0.10$ as possible counterparts.

We uncover PACS secure or possible
counterparts for 22 SMGs, corresponding to a PACS identification rate of $39\%$ of our whole SMG sample, see also Table~\ref{tab:cptproperties} for details. Our PACS identification rate of $39\%$ is
lower than the  rate of $54\%$ found by \citet{mag10}. The reason is that \citet{mag10} focus on already radio-identified SCUBA and AzTEC sources with spectroscopic redshifts mainly. The top panel of Fig.~\ref{fig:pacsdetection} displays the (sub)millimeter-PACS separation of our association. We did not find evidence of systematic offsets. We find PACS counterparts at a mean (sub)millimeter-PACS positional offset of $\Delta_{(submm)mm-PACS100}$$=5.0\arcsec\pm2.1$\arcsec\/ and
$\Delta_{(submm)mm-PACS160}$$=4.9\arcsec\pm2.0$\arcsec. This is consistent with a typical submm/mm position error of about 3\arcsec\/ to 5\arcsec\/ and strengthens our choice of the FWHM of the bolometric data as search circle for PACS counterparts. The top panel of Fig.~\ref{fig:pacsdetection} shows clearly a dominance of secure counterparts. Given this statistical approach, associations with offsets much larger than the average might not always be correct.
%
\begin{table}[!h]
\caption{PACS associations of 56 Submillimeter Galaxies in GOODS-N}             
\label{tab:cptproperties}      
\centering                          
\begin{tabular}{l c c}        
\hline\hline                 
 & PACS~$100~\mu$m & PACS~$160~\mu$m  \\    
\hline                        
\#\/ secure association &12&15\\
\#\/ possible association &4&5\\
total identification rate&28.6\%\/&35.7\%\/\\
(sub)mm-PACS offset&$5.0\arcsec\pm2.1\arcsec$&$4.9\arcsec\pm2.0\arcsec$\\\\
&\multicolumn{2}{c}{PACS~$100+160~\mu$m}\\
\#\/ total association of SMGs&\multicolumn{2}{c}{22 (39.3\%\/)}\\
\#\/ only detected at PACS~$160~\mu$m&\multicolumn{2}{c}{6}\\
\#\/ only detected at PACS~$100~\mu$m&\multicolumn{2}{c}{2}\\
new CPTs identified by PACS&\multicolumn{2}{c}{1}\\
\#\/ PACS and radio blank fields&\multicolumn{2}{c}{12 (21.4\%\/)}\\
\hline                                   
\end{tabular}
\end{table}
%

We detect 20 (16) SMGs at 160 (100) ~$\mu$m and classify 15 (12) of
them as secure and 5 (4) as possible counterparts. The number of PACS
counterparts at 160~$\mu$m is higher than at 100~$\mu$m.  This is
expected as the 160~$\mu$m measurements lie close to the FIR peak.  We
note that based on the corrected Poissonian probability {\it p} each
PACS detection within the bolometer beam (our search circle) is
classified as associated SMG counterpart. Typical PACS fluxes of these
dust-obscured high-z sources range between 4.0~mJy to 34.9~mJy at
100~$\mu$m and 5.0~mJy to 65.0~mJy at 160~$\mu$m.

We have 43 (28 are secure; 15 are possible) radio-identified SMGs in
our sample, about 50\% of them are seen at PACS wavelengths. Vice-versa,
only one PACS association (secure) is undetected at 1.4~GHz, see
detailed discussion of GN13 (alias HDF850.4) at end of
Section~\ref{sec:diagnostic}. One-third of PACS detected SMGs have at
least two VLA counterparts. 

\section{Diagnostic Potential of PACS observations of Submillimeter Galaxies}
\label{sec:diagnostic}
About twice as many SMGs are identified by the VLA than by PACS. VLA
observations are more sensitive for sources at redshifts up to
$z=4$. None of the well-known, spectroscopically identified SMGs at
$z=4$ --- GN20, GN20.2a/b and GN10
\citep{dad09a,dad09b,wan07,wan09,dan08} --- have been significantly
detected in our PACS imaging, see middle panel in
Fig.~\ref{fig:pacsdetection}. Another highly promising $z\geq4$
candidate, HDF850.1 \citep[][]{hug98,dun04,cow09} is also not seen by
PACS.  However, all these SMGs have radio counterparts. Focusing on
our spectroscopic subsample of 15 sources, SMG$_{spec}$, we find a
trend that the fraction of SMGs detected at PACS bands decreases with
redshift (see middle panel in Fig.~\ref{fig:pacsdetection}), being
explained by the fact that PACS fluxes drop with increasing
redshift. No source beyond $z=2.00$ (GN06) at 100~$\mu$m and $z=2.58$
(GN04) at 160~$\mu$m is significantly detected by PACS. However, we do
not find evidence that the PACS detection rate of SMGs correlates with
the SCUBA flux density (Fig.~\ref{fig:pacsdetection}). In addition,
VLA observations provide subarcsecond accurate positions which are
essential for the proper identification in the optical and
near-infrared bands and follow-up spectroscopic observations. PACS
cannot deliver positions at these accuracies.  To summarize, ultra-deep VLA
observations still remain the best and most effective approach to
identify SMG counterparts, even up to redshift $z=4$.

\begin{figure}[!b]
   \centering
   \includegraphics[width=9cm]{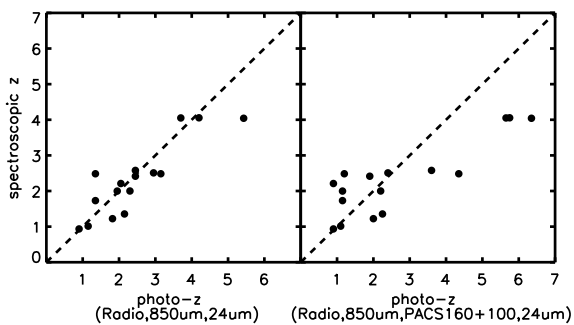}
      \caption{Comparison of spectroscopic and photometric redshifts. In the left panel  we show photometric redshifts based on radio,  850~$\mu$m and 24~$\mu$m flux densities measurements for SMG counterparts with spectroscopic redshifts in GOODS North \citep[adopted from][]{dad09a}. There is a fairly good agreement. In the right panel, we show photometric redshift estimates including PACS measurements which are less reliable.}
         \label{fig:photoz}
   \end{figure}

Using only radio, 850~$\mu$m and 24~$\mu$m flux densities,
\citet{dad09a} derived accurate ($\Delta z/(1+z)\sim0.1$) radio-IR
photometric redshifts for SMGs. Motivated by this work, we apply the
radio-IR photometric redshift code presented in \citet{dad09a} and
calculate photometric redshifts by adding our PACS measurements to
existing SCUBA, MIPS~24~$\mu$m and VLA 1.4~GHz flux measurements of
our SMG sample. Naively we would have expected to improve the accuracy
of the radio-IR photometric redshifts by adding FIR measurements
shortwards to the IR dust peak. In Fig.~\ref{fig:photoz} we show the
results of our attempt using template libraries of local galaxies
\citep{cha01}. In the left panel we show photometric redshifts
\citep[adapted from][]{dad09a} based on radio, MIR and submm
observations for SCUBA counterparts with known spectroscopic redshifts
in GOODS North which agree fairly well. In the right panel we show
photometric redshift estimates adding the PACS measurements to our
previous measurements in the radio, MIR and submm. Clearly the
obtained accuracy decreases and these photo-z estimates are poorer. Omitting certain measurements (e.g., MIPS~24$\mu$m and/or
VLA~1.4~GHz) gives even worse results. We obtain similar results by
using template libraries from \citet{dal02} and the average SED
constructed by \citet{mic10} of 73 spectroscopically identified SMGs
\citep{cha05}. This result means that most likely the far-infrared
properties of SMGs are different from templates built to describe
local galaxies. These observed differences arise most probably
from the fact that SMGs are colder than local ULIRGs of same
luminosity \citep{cha05,mag10,elb10,chan10}. Further investigations
are needed to fully exploit the Herschel-PACS imaging in order to
obtain accurate photometric redshifts.
%

\begin{figure}[!t]
   \centering
   \includegraphics[width=8.5cm]{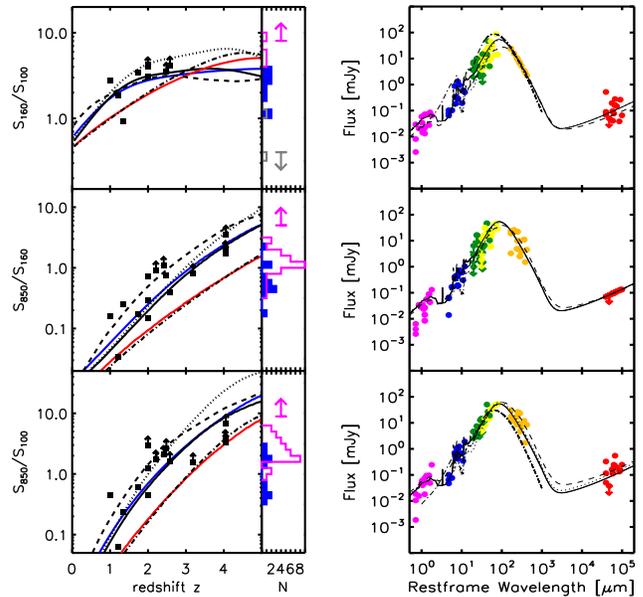}
      \caption{{\it Left Panels}: From top to bottom we plot the flux ratios $S_{160~\mu m}/S_{100~\mu m}$, $S_{850~\mu m}/S_{160~\mu m}$ and $S_{850~\mu m}/S_{100~\mu m}$ versus redshift for SMGs with spectroscopic redshifts. Template SEDs  of a local LIRG with $L_{IR}\approx~1\times 10^{11}~L_{\sun}$ (blue-solid line), a local ULIRG with $L_{IR}\approx~1\times 10^{12}~L_{\sun}$ (black-solid line) and a HyLIRG with  $L_{IR}\approx~1\times 10^{13}~L_{\sun}$ (red-solid line), all from \citet{cha01}, and an average SMG \citep[black-dashed line,][]{mic10}, are shown for comparison. Average SEDs including Spitzer-IRS spectroscopy of local ULIRGs \citep[black-dashed-dotted line,][]{arm07} and SMGs \citep[black-dotted line,][]{pop08} are shown as well. The histograms present the distribution of SMGs without spectroscopic redshifts (blue-filled), lower limits of flux ratios are displayed by the pink-empty histogram. For the  $S_{160~\mu m}/S_{100~\mu m}$ flux ratio we show the upper limit as gray-empty histogram. The pink (gray) arrow underlines the limits. {\it Right Panels:} From top to bottom, we plot the composite rest-frame SED --- IRAC~3.6~$\mu$m (pink), MIPS~24~$\mu$m (blue), PACS 100~$\mu$m (green), PACS 160~$\mu$m (green), SCUBA 850~$\mu$m (gold), VLA 20~cm (red) --- of our spectroscopic SMG sample normalized on SCUBA, VLA and PACS 160~$\mu$m fluxes. Template SEDs  of ULIRGs and SMGs are shown for comparison. Both data points and templates are normalized on an SMG with a SCUBA flux of 8~mJy at $z=2.2$.}
\label{fig:pacscolors}
\end{figure}

We already know the spectroscopic redshift for 15 SMGs in our
sample. However, in absence of spectroscopic redshifts for most of our
sources, searching for correlations between observed properties is
essential for constraining the nature, redshift and evolution of
SMGs. We explore if flux ratios involving PACS 100~$\mu$m and
160~$\mu$m measurements can be used as a rough redshift
indicator. Fig.~\ref{fig:pacscolors} displays the results of our
analysis. The PACS color ($S_{160~\mu m}/S_{100~\mu m}$) of SMGs does
not vary with redshift and is consistent with the prediction of models
and observed SEDs from \citet{cha01}, \citet{mic10}, \citet{arm07} and
\citet{pop08}. The PACS colors seem to be slightly redder than
predicted by the various models and may indicate a difference between
the observed FIR-SED of SMGs and templates describing fairly well
local infrared galaxies. The composite rest-frame SED for the
SMG$_{spec}$ (Fig.~\ref{fig:pacscolors}) subsample shows their
diversity in the far-infrared.  Their far-infrared colors are
different from the template SEDs, most notably for the local luminous
infrared galaxies. This may explain our finding reported above that we
are not able to obtain accurate IR-radio photometric redshifts by
including PACS measurements.

The PACS fluxes are redshift dependent, whereas the (sub)millimeter
flux density at $z\ge1$ is not \citep[due to the negative
K-correction; e.g.,][]{bla02}, thus one would expect to see a trend
between (sub)mm/PACS flux ratio for both PACS bands (see also
the predictions based on different templates in Fig.~\ref{fig:pacscolors}). Indeed, we find a fairly
clear trend both for $S_{850~\mu m}/S_{160~\mu m}$ respectively
$S_{850~\mu m}/S_{100~\mu m}$ versus spectroscopic redshift for the
SMG$_{spec}$ subsample\footnote{For SCUBA sources from \citet{pop06}
we use the deboosted flux. The same we do for AzTEC and MAMBO
sources. We converted the AzTEC and MAMBO mm flux densities into SCUBA
flux densities by assuming a flux ratio of $S_{850}/S_{1~mm}$ of 2.5
for a ULIRG-like SED at redshift $z\sim2-3$.} over the whole redshift
range spanned by our SMGs from $z~\sim1 - 4$. We conclude that the
(sub)mm/PACS flux ratio seems to be a useful albeit crude
redshift indicator and may help to select/mark SMGs at very
high-redshifts. We explored also VLA/PACS and PACS/MIPS~24~$\mu$m flux
ratios versus redshift but did not observe any correlation, being
consistent with the fact that VLA, MIPS~24~$\mu$m and PACS fluxes drop
with increasing redshift. Based on this finding, we investigate the
(sub)mm/PACS ratio for the remaining 41 SMGs without any reliable
redshift. This analysis shows that PACS-detected SMGs tend to have
lower flux ratios than the PACS-undetected SMG sample (Fig.~\ref{fig:pacscolors}). This trend seems
to be more prominent for the $S_{850~\mu m}/S_{160~\mu m}$ flux
density ratio. Our analysis may indicate that indeed as previously
discussed PACS-undetected SMGs could tend to lie at higher redshifts
and the (sub)mm/PACS flux ratio could be a crucial tool to select the
very high-z tail of the SMG population.

%

The previous analysis bridges with the discovery that 12 out of 56
SMGs ($21\%\/$) in our sample are blank fields both at PACS and radio
wavelengths. The nature of these sources is still not clear. These
SMGs could be either spurious sources which is mainly assumed, or at
redshifts higher than $z>4$. Indeed, how reliable are these (sub)mm
sources detections? The sample consists of seven submillimeter and five
millimeter selected sources (of the latter population, one is seen by
AzTEC and four by MAMBO). We checked each source for its
reliability. The mm sources are selected with SNR $\geq4$. Four of the
seven SCUBA galaxies are listed in both \citet{pop06} and
\citet{wan04}. The fact that \citet{pop06} and \citet{wan04} used
(nearly) the same SCUBA data, but performed independent data reduction
and source extraction, is giving more weight to the reliability of
these detections. As a sanity check we inspected the SCUBA and MAMBO
galaxies in the AzTEC 1~mm map. Our analysis suggests that most of
these sources are reliable. One of them has a SNR $\sim9$ at
850~$\mu$m.  However, we note that only mm interferometric
observations will reveal unambiguously the reliability of these
sources.  Their non-detection both at radio and PACS wavebands in
combination with its (sub)mm/PACS flux ratio may suggest that these
sources could lie at higher redshifts than the typical SMG population
(e.g., $z>3-4$).
   \begin{figure}[!t]
   \centering \includegraphics[width=6cm]{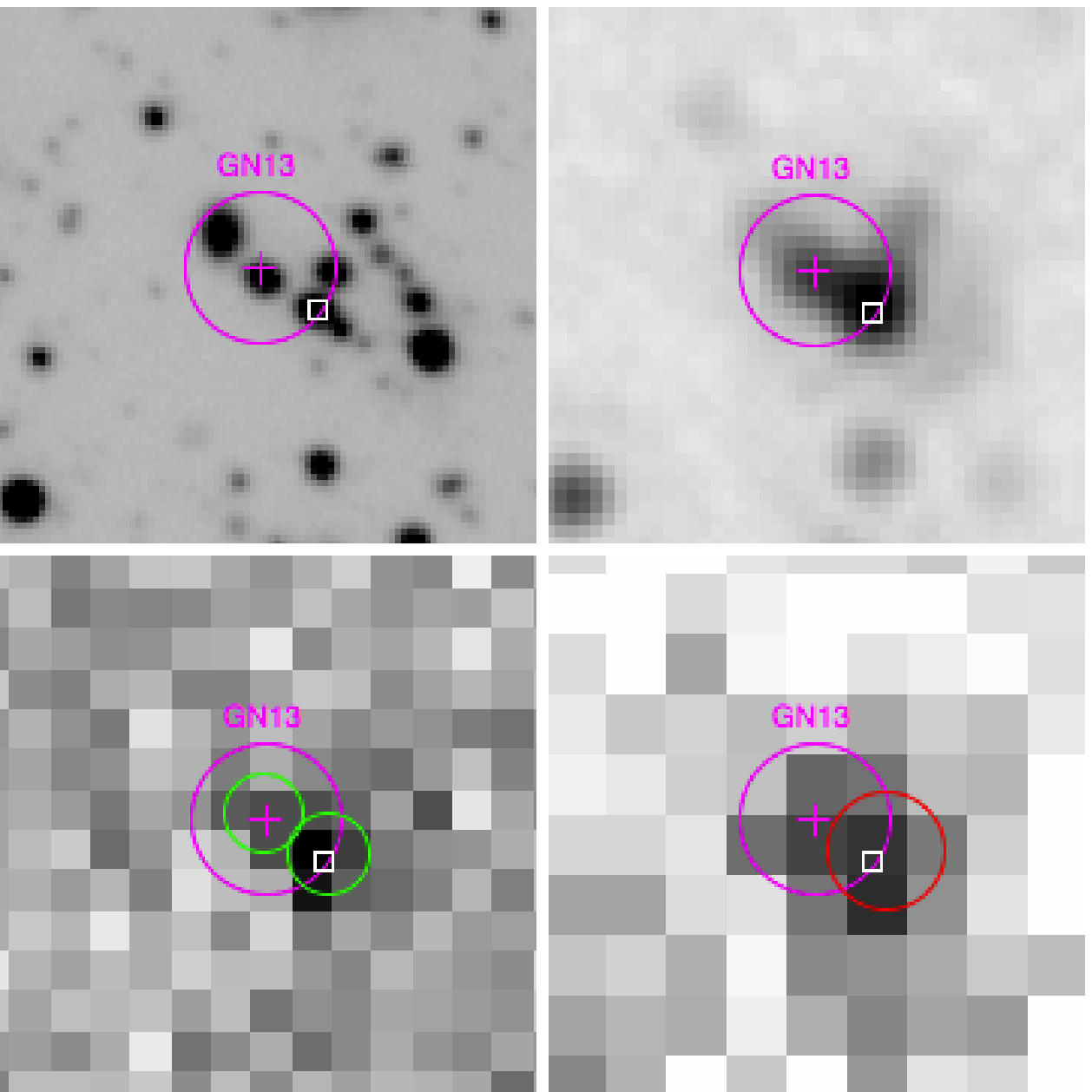}
      \caption{From left to right and top to bottom we show the
      $50\arcsec\/\times50$\arcsec\/ IRAC~3.6$\mu$m, MIPS~$24~\mu$m,
      PACS~100$\mu$m and PACS~160$\mu$m images of the field of GN13
      alias HDF850.4 \citep{pop06,hug98}, north is up and east to the
      left. The 7.5\arcsec\/ search radius circle at the nominal
      SCUBA position of GN13 (magenta cross) is drawn. Green (red)
      circles represent PACS~100$\mu$m (PACS~160$\mu$m) $\ge3\sigma$
      detections and the white square a VLA source.  \citet{dad09a}
      suggested that the counterpart at $z=0.457$ in South-West
      direction $\sim7$\arcsec\/ away from GN13 is a
      mis-identification. We propose the PACS~100~$\mu$m detection on
      top of the nominal SCUBA position (magenta cross) as new
      counterpart at $z\sim1-2.5$.}
         \label{fig:gn13}
   \end{figure}
%

We conclude this work by presenting the identification of an SMG
counterpart having a PACS but no VLA detection. \citet{dad09a} suspected
that the source at $z_{spec}=0.457$, about 7\arcsec\/ away from the
SCUBA position, could not be the correct counterpart of GN13 alias
HDF850.4 \citep{pop06,hug98}. We detect a faint ($4.2\sigma$), secure
counterpart at 100~$\mu$m, being very close to the bolometer position
($\sim0.6\arcsec$). Its reliability is fortified by a 200~$\mu$Jy
strong, secure counterpart at 24~$\mu$m. Using the flux ratios
$S_{850~\mu m}/S_{100~\mu m}=0.38$ and $S_{850~\mu m}/S_{160~\mu
m}>0.37$ as a coarse redshift indicator suggests a redshift of
$z\sim1-2.5$.

\begin{acknowledgements}
PACS has been developed by a consortium of institutes led by MPE (Germany) and 
including UVIE (Austria); KUL, CSL, IMEC (Belgium); CEA, OAMP (France); MPIA 
(Germany); IFSI, OAP/OAT, OAA/CAISMI, LENS, SISSA (Italy); IAC (Spain). This 
development has been supported by the funding agencies BMVIT (Austria), 
ESA-PRODEX (Belgium), CEA/CNES (France),
DLR (Germany), ASI (Italy), and CICYT/MCYT (Spain). We thank Alex Pope and Lee Armus for providing SEDs of SMGs and local ULIRGs. We thank an anonymous referee for useful comments that improved our manuscript. We acknowledge the funding support of  the ERC-StG grant UPGAL 240039
and ANR-08-JCJC-0008.
\end{acknowledgements}

\end{document}